# Noninvasive three-dimensional optoacoustic localization microangiography of deep tissues

Xosé Luís Deán-Ben, Justine Robin, Ruiqing Ni, and Daniel Razansky

*Institute for Biomedical Engineering and Institute of Pharmacology and Toxicology*
*University of Zurich and ETH Zurich, Switzerland*

**ABSTRACT**

Structural microvascular alterations and dysfunction serve as key disease indicators of cancer, diabetes, ischemic stroke, neurodegenerative disorders and many other conditions. *In vivo* visualization of the microvasculature has traditionally been restricted to millimeter scale depths accessible with optical microscopy. Optoacoustic imaging has enabled breaking through the barrier imposed by light diffusion to map functional hemodynamic parameters in deep-seated vessels, but diffraction and dispersion of ultrasound waves in heterogeneous living tissues prevents reaching capillary resolution. Herein, we demonstrate three-dimensional microangiography of deep mouse brain beyond the acoustic diffraction limit (<20 μm resolution) through the intact scalp and skull via optoacoustic localization of sparsely-distributed highly absorbing microparticles. This was enabled by devising 5 μm sized extremely absorbing dichloromethane microdroplets exhibiting four orders of magnitude higher optical absorption than red blood cells at near-infrared wavelengths, thus facilitating *in vivo* compatibility and single particle sensitivity in the presence of highly absorbing blood background. Accurate mapping of the blood flow velocity within microvascular structures was also facilitated by the high 3D frame rate of the optoacoustic tomography system. We further show that the detected optoacoustic signal intensities from the localized particles may serve for estimating the light fluence distribution within optically heterogeneous tissues, a long-standing quantification challenge in biomedical optics. Given the intrinsic sensitivity of optoacoustics to various functional, metabolic and molecular events in living tissues, this new approach paves the way for non-invasive deep-tissue microscopic observations with unrivaled resolution, contrast and speed.

**Introduction**

New methods enabling breaking through established resolution barriers have massively impacted life sciences ever since first optical microscopes have overcome the resolution limits of a naked eye [1]. Optical microscopy performance was subsequently enhanced to reach the light diffraction barrier at imaging depths not affected by photon diffusion - typically <1 mm within animal models [2,3]. Higher resolution has been achieved with electron microscopy [4], while a myriad of technologies based on x-rays, ultrasound (US) or nuclear magnetic resonance have been developed for deep tissue imaging [5]. Yet, the unique ability of optical methods to sense specific molecules *in vivo* often makes them the preferred choice for biological observations. Recent efforts have thus been directed toward overcoming light diffraction barriers in fluorescence microscopy [6] and developing new optical imaging approaches operating at mesoscopic and macroscopic depths [7,8].

Optoacoustic (OA, photoacoustic) imaging has further emerged as a hybrid modality synergistically combining optical contrast and US resolution to provide otherwise unavailable functional and molecular information from deep tissues [9,10]. A number of advanced OA embodiments enabled breaking through the light penetration barrier and attained high spatial resolution (depth-to-resolution ratio of ~200) at millimeter to centimeter scale depths [11,12], while further providing unprecedented 3D imaging speeds in the kilohertz range [13]. These unique features facilitate the interrogation of biological samples with very versatile optical-absorption-based contrast, as well as spatial and temporal scales not accessible by other modalities. OA microvascular imaging is yet restricted to shallow regions e.g. superficial cortical layers [14], while individual capillaries can only be visualized up to a depth of ~1 mm, where light can be efficiently focused [15,16]. Much like for pulse-echo US, the resolution of OA imaging in the light diffusion regime is affected by frequency-dependent acoustic attenuation, which effectively reduces the bandwidth of the collected signals and establishes a depth-dependent acoustic diffraction limit. Recently, ultrafast US broke through this barrier via localization of single circulating microbubbles[17]. Accurate particle localization is however hampered by acoustic distortions produced when pressure waves propagate through strongly mismatched tissues such as the skull bone in brain imaging applications. In this regard, OA has an advantage over pulse-echo US as it only involves unidirectional propagation of waves generated by light absorption within the sample. We previously demonstrated the basic feasibility of localization OA tomography (LOT) by visualizing the flow of absorbing microspheres in scattering phantoms [18]. LOT was shown to significantly improve the spatial resolution of OA tomography (OAT) and microscopy [18-20], while further enhancing the visibility of structures under limited-view tomographic conditions. a major deficiency of most OA systems [21]. Combined with particle tracking, LOT can additionally provide dynamic information on the blood flow, namely blood velocity, currently unattainable with OAT. *In vivo* LOT is however hampered by the strong background absorption of blood. Red blood cells (RBCs) consist of ~270·10$^6$ hemoglobin molecules densely distributed in blood (~50% v/v), while typical OA contrast agents, e.g. based on small molecules or nanoparticles, have significantly lower absorbance on a per-unit basis [22]. Individual liquid droplets and circulating tumor cells could be localized in the mouse brain after intracardiac injection, but such large (~30 μm diameter) bodies are likely to be arrested in the capillary network [23,24], thus hindering *in vivo* compatibility. Also important is the fact that vascular structures arbitrarily oriented in all three dimensions at different depths cannot be efficiently captured with cross-sectional (two-dimensional, 2D) OAT systems, thus three-dimensional (3D) tomographic imaging turns essential for an optimal LOT performance.

Herein, we demonstrate noninvasive 3D OA localization microangiography in the deep mouse brain by means of intravenous (*i.v.*) bolus injection of extremely absorbing dichloromethane (DCM) microdroplets with sizes in the order of RBCs. The small size of the droplets prevents capillary blockage, while their strong absorption facilitates individual detection and accurate localization within the field of view (FOV) of a volumetric OAT system with a single shot excitation.

**Results**

The basic principle of LOT along with a lay-out of the tomographic approach used to image the mouse brain is depicted in Fig. 1a. A short-pulsed (nanosecond duration) laser tuned to an optical wavelength of 780 nm was used to simultaneously illuminate the entire brain and generate pressure (US) waves via thermal expansion that are eventually detected with a spherical array of transducers (see methods for a detailed description). These pressure waves are produced by endogenous absorbers, mainly hemoglobin in RBC, as well as by the exogenously administered absorbing microdroplets required for LOT. This leads to a static background OA signal corresponding to vascular structures of different sizes superimposed onto a dynamically changing signal generated at the particles flowing in blood. LOT can only be performed if the signal generated by an individual (isolated) particle is sufficiently high to be detected in a single image voxel. The background signal per voxel is proportional to the number of RBCs enclosed, which can reach 100s to 1000s for the given 100 µm resolution of the imaging system. The DCM droplets employed have an approximate concentration of ~200 mM of IR-780 dye (extinction coefficient ~250·$10^3$ $M^{-1}cm^{-1}$). As a reference, the extinction coefficient of hemoglobin at 780 nm and concentration in RBC are ~$10^3$ $M^{-1}cm^{-1}$ and ~5 mM, i.e., the total light absorption in the DCM droplet is ~$10^4$ times higher than in a RBC. The average diameter of the droplets (Fig. 1b, 5.5 µm) is smaller than the disk diameter of murine RBCs (4-7 µm) [25]. Individual droplets could clearly be detected in the OA images after clutter removal with singular value decomposition (SVD) filtering (see methods for a detailed description). Co-registration of OA images with the Allen mouse brain atlas [26] (Fig. 1c) further served as anatomical reference to identify the droplets' position at a given time point. For example, three microdroplets in the anterior cerebral artery system (M1, azygos pericallosal artery and M2, middle internal frontal artery penetrating branches) and in the pial surface (M3) are clearly visible in Fig. 1c. A continuous flow of particles was observed over the entire image sequence, indicating that no capillary arrest was induced. Note that the OAT system's ability to cover an entire 3D region with single shot excitation is of particular importance for an accurate trajectory estimation, e.g. when a droplet flows from a deep-seated artery to the superficial cortical vasculature (Fig. 1d). In addition, the high temporal resolution is paramount for accurately quantifying the relatively high blood flow velocity.

LOT images were subsequently rendered by aggregating the localized positions of the microdroplets flowing through the vascular network over a temporal sequence of OA images (see methods for details). No average or compounding of OA frames was required for localizing individual particles. This is extremely important considering the high blood flow velocity in the mouse brain (ca. 200 ml/100g/min, depending on anesthesia and depth from the surface [27,28]). The effective integration time of the images used for particle localization must be minimized. OA images acquired with a single laser pulse (single shot excitation) are very well defined in time by the pulse duration, which facilitates accurate localization of particles moving with high velocity. The enhanced resolution achieved with LOT with respect to standard (acoustic resolution) OAT is clearly observed e.g. in the dorsal view of the mouse brain vasculature (Figs. 2a,b). Pial arterioles such as the anterior cerebral artery, the middle cerebral artery and the posterior cerebral artery, as well as the superior sagittal sinus are clearly identified in the LOT image. Pial arterioles and penetrating arterioles (long and short cortical branches) across the cortex over the expected thickness (~1 mm) [14,29] are also clearly visible (approx. Bregma -0.22 mm to -1.22 mm, Fig. 2b coronal view). Such features cannot be resolved in standard OAT images not only because of its inferior resolution but chiefly due to limited-view effects hindering reconstruction of vertically-oriented structures [21]. Vascular structures separated by 22 µm could be resolved over a depth range of ~1.5 mm from the brain cortex (~2.5 mm from the skin surface) through the intact scalp (estimated thickness 0.5 mm) and skull (estimated thickness 0.4-0.6 mm) [30,31]. Note that the spatial resolution in LOT depends on the number of droplets passing through each reconstructed voxel (10x10x10 µm$^3$) and thus could potentially be improved by using a longer acquisition time or a higher droplet concentration – approximately 20 droplets were localized in each volumetric frame in the present case.

The high temporal resolution of the 3D OAT imaging system further enabled estimating the blood flow velocity via particle tracking (see methods section for a detailed description). The reconstructed velocity map (Fig. 2d) is consistent with previously reported values for the mouse brain, i.e. from several cm/s in major cerebral arteries - typical blood flow range is 0.5-1.6 ml/g/min depending on the brain region and depth [32,33] - down to below 5 mm/s in smaller vessels [34,35]. It was further possible to accurately assess the velocity profile across relatively large vessels, revealing a Poiseuille-like flow profile (inlet in Fig. 2d). The capability to accurately quantify the blood flow represents an important new feature in the unique functional imaging portfolio of OAT, which is already capable of measuring various hemodynamic parameters such as blood volume and oxygen saturation [36,37]. Finally, it is important to take into account that the OA signal generated by the microdroplets is directly proportional to the local light fluence. We observed a clear signal decay with depth (Fig. 2e, top, approx. Bregma -1.46 mm). The light fluence distribution could then be estimated by fitting an exponentially decaying function to the measured signal intensities at different positions in this section (Fig. 2e bottom, see methods for a detailed description). As the tissue composition is different in white/gray matter and across brain regions, a heterogeneous fluence distribution is generally expected. Brain atlas co-registration was further facilitated by identifying vessels in the coronal section, e.g. the pial artery, transverse hippocampal artery/vein and middle cerebral artery, thus accurately assessing which brain regions are efficiently illuminated (Fig. 2e bottom).

**Discussion**

The ability to localize and track intravenously injected micron-size particles in 3D and in the presence of highly absorbing blood background empowers LOT with an unprecedented capacity for microscopic *in vivo* imaging of deep optically opaque tissues. Breaking through the acoustic diffraction barrier facilitates volumetric visualization of microvascular structures not captured in standard OAT images. It was shown that LOT can achieve 20 µm resolution across >3 mm depth range through an intact scalp and skull. Apart from its superb spatial resolution and better visibility of structures under limited-view tomographic conditions, LOT enables quantification of the local light fluence and blood flow velocity. Key to these enabling features are the high 3D imaging frame rate (100 Hz), short duration of the excitation laser pulse (<10 ns) and sufficient detection sensitivity for capturing entire 3D image volumes with single-shot excitation.

The use of strong micron-size absorbers is an important pre-requisite for successful *in vivo* application of the LOT technique as they effectively break the continuity in the absorption distribution induced by densely-packed RBCs and can easily be localized, even without employing broad angular coverage for OAT image acquisition. Indeed, LOT is capable of rendering accurate vascular images under limited-view conditions, thus overcoming another major limitation of the commonly employed OAT systems [21]. Limited-view effects are associated with the speckle-free nature of OAT and have been commonly averted by artificially inducing speckle grains in the images [38-40], which is equally possible in LOT even if the injected particles cannot be individually distinguished. The ability of LOT to directly measure light fluence distribution in deep tissues and anatomical co-registration based on atlas delineation of white/gray matter structures, addresses another important long-standing challenge in biomedical optics [41]. Particularly, OA image normalization with fluence is essential for achieving accurate readings of the concentration of agents or rendering physiologically-relevant blood oxygenation measurements in the presence of spectral coloring effects in deep tissues [42,43]. Fluence estimation may further be facilitated with monodisperse droplets, which can potentially be synthetized with microfluidic chips.

The microdroplets employed in this work have a similar size to food and drug administration (FDA)-approved microbubbles used as US contrast agents. The demonstrated results may thus foster clinical translation of OA imaging. On the other hand, the FDA-approved indocyanine green (ICG) agent has a similar extinction coefficient as the IR780 dye used in our study, making it an ideal candidate for undergoing the regulatory approval process once properly encapsulated into a microparticulate agent. Beyond their relevance for LOT, strongly absorbing microparticles smaller than RBCs may emerge as powerful agents for enhancing OA contrast significantly outperforming the existing OA contrast agents based on small molecule dyes and nanoparticles.

LOT can greatly impact our understanding of brain microvascular organization and function and provide new insights into major research challenges in neuroscience [44]. It can enhance the capabilities of OAT to provide hemodynamic readings e.g. of oxygen saturation or cerebral blood volume by additionally enabling quantifying cerebral blood flow or diameter changes in microvascular structures. LOT may then emerge as a unique tool for studying neurovascular coupling and stimulus responses e.g. in subcortical regions not accessible with other modalities [45]. Cerebrovascular structural alterations and dysfunction play critical roles and are key disease biomarkers in multiple diseases such as hypertension, ischemic stroke, brain cancer, traumatic brain injury or Alzheimer's and tauopathy diseases [46-48]. Magnetic resonance imaging (MRI) and x-ray computed tomography (CT) have been used to detect vessel remodeling, blood flow disturbances and vascular density in humans and in murine disease models with suboptimal resolution [49]. Two-photon microscopy and laser speckle imaging have alternatively been used to map the microvascular blood flow, but are only applicable at shallow depths (<1 mm from the accessible surface) [27,32,50]. The ability of LOT to visualize blood flow dynamics in the middle and anterior cerebral arteries (both pial and penetrating arterioles) enables e.g. studying the mechanisms of reperfusion after middle cerebral artery occlusion [51], monitoring treatments for neuroprotection neuroplasticity after stroke or characterizing intracerebral hemorrhages after immunotherapy [52].

Taken together, these unique enabling features can massively impact our understanding on the structural and functional properties of cerebral microvasculature under physiological and diseased conditions. In a broader perspective, LOT can be used to facilitate early diagnosis based on bio-markers associated with microcirculatory alterations in diabetes, cancer, cardiovascular disorders, ischemic stroke or neurodegenerative diseases, while additionally providing new insights into disease progression, efficacy of drugs and other therapeutic interventions. In conclusion, the newly developed capacity for rapid volumetric mapping and characterization of microvascular structures *in vivo* with spatial resolution beyond the acoustic diffraction barrier is poised to provide unprecedented insights into the anatomy and function of optically opaque organisms. We anticipate that the advantages provided by LOT, combined with the unique features of state-of-the-art OAT, will massively impact a large number of studies on clinically relevant diseases and further foster the growing use of OA as a biomedical imaging tool.

**Methods**

**Microdroplet synthesis and characterization**

A suspension of dichloromethane (DCM, Sigma Aldrich, 270997) microdroplets in water was produced following a standard emulsification procedure. The disperse (DCM) phase was prepared by diluting 30 g of IR-780 iodide (Sigma-Aldrich, 425311) in 200 ml of DCM to achieve a dye concentration of approximately 200 mM. The continuous (water) phase was prepared by adding 3% (v/v) Tween 20 surfactant (Sigma Aldrich, P1379) in phosphate buffered saline (PBS, Sigma Aldrich, 79378). 50 µl of disperse phase and 1.5 ml of continuous phase were mixed in a 2 ml Eppendorf and vortexed at speed 9 during 30 s. The continuous phase was cooled down to approximately 4°C before vortexing to avoid vaporization of DCM. The resulting emulsion was eventually filtered with a cell strainer with pore size 10 µm (PluriStrainer 10 µm, pluriSelect Life Science, Leipzig, Germany) and subsequently inspected in a stereo microscope (Carl Zeiss AG, Oberkochen, Germany) to verify that no particles with irregular (non-spherical) shape, arguably corresponding to the dye powder, are present in the suspension. The diameters of the droplets obtained after filtering the suspension with the cell strainer with pore size

10 μm were measured in the images taken with a bright field microscope (Leica Camera AG, Wetzlar, Germany, 5X objective). Specifically, a total of 500 circles were automatically detected and characterized (MATLAB function imfindcircles) and the results were displayed as histograms (Fig. 1b).

**Optoacoustic imaging system**

OA imaging of the mouse brain was performed with a tomographic imaging system schematically illustrated in Fig. 1a. A spherical array of 512 US sensing elements (Imasonic SaS, Voray, France) was used to collect the corresponding OA signals generated via excitation with an optical parametric oscillator (OPO)-based short-pulsed (<10 ns) laser (Innolas GmbH, Krailling, Germany) tuned to 780 nm, corresponding to the peak absorption of the dye. The elements of the US array have 7 MHz central frequency and >80% detection bandwidth. The array features a central aperture with 8 mm diameter and 3 lateral apertures with 4 mm diameter located at 45° elevation angle and equally spaced (120°) in the azimuthal direction. Light was guided with a custom-made 4-arm fiber bundle (CeramOptec GmbH, Bonn, Germany) through the apertures of the array. This provided an approximately uniform illumination profile on the mouse brain surface with optical fluence <20 mJ/cm$^2$. The OA signals were digitized at 40 megasamples per second with a custom-made data acquisition system (DAQ, Falkenstein Mikrosysteme GmbH, Taufkirchen, Germany) triggered with the Q-switch output of the laser and transmitted to a computer via Ethernet.

**Animal model**

Female athymic nude mice (n=5, 6-8 weeks old, Janvier Lab, France) were used for *in vivo* experiments. Animals were housed in ventilated cages inside a temperature-controlled room under a 12-hour dark/light cycle. Pelleted food and water were provided *ad-libitum*. All experiments were performed in accordance with the Swiss Federal Act on Animal Protection and were approved by the Cantonal Veterinary Office Zürich (ZH 161/18).

***In vivo* LOT imaging**

Mice were anaesthetized with isoflurane (4% v/v for induction and 1.5% during the experiments, Abbott, Cham, Switzerland) in an oxygen/air mixture (100/400 mL/min). OA imaging of the brain region was performed with the head of the mouse fixed into a custom-designed stereotactic mouse head holder coupled to a breathing mask (Narishige International, Japan). Blood oxygen saturation, heart rate and body temperature were continuously monitored (PhysioSuite, Kent Scientific) and the temperature was maintained at ~36°C with a heating pad. Bolus injection (*i.v.*) of 100 ml of the microdroplet emulsion was performed 30 s after the beginning of data acquisition, where the total acquisition time was 420 s. Four mice were euthanized under deep anesthesia (5% isoflurane for 5 minutes) and subsequently decapitated without waking them up. To examine the possible toxicity of the microdroplet emulsion, the health condition and behavior of one mouse was monitored (3 times/day) for 24 hours after the *in vivo* imaging. The mouse was subsequently euthanized as described above. The mouse fully recovered and no effect on the well-being and behavior was observed.

**Image reconstruction**

OA images for a volume of 10x10x5 mm$^3$ (100x100x50 voxels) were reconstructed with a graphics processing unit (GPU)-based implementation of a back-projection formula [53]. Prior to reconstruction, the collected signals were band-pass filtered with cut-off frequencies 0.1 and 9 MHz. A singular value decomposition (SVD) clutter filter was further applied to the raw OA signal data to reconstruct the OA images used for single droplet localization [54]. This enabled isolating the signal fluctuations ascribed to absorbing particles flowing in the vasculature from the static (background) signal coming from endogenous absorbers such as hemoglobin (20 singular vectors out of 500 were removed). A reference OA image was also reconstructed after applying the same filter for comparison purposes (Fig. 2a).

**Image registration**

Registration between OA images and Allen mouse brain atlas was performed to further provide a better anatomical reference for identifying cerebral vascular structures. Specifically, an annotated Allen brain atlas [26] was used to identify brain regions in the OAT and LOT datasets. The reference atlas was manually aligned with the OA images (Amira) and for annotation of the vessels [55], which provided the best contrast for distinguishing the cerebral vasculature.

**Droplets localization and tracking**

Isolated microdroplets are strong absorbers smaller in size than the resolution of our imaging system and thus appear in the image as the local point spread function (PSF). In each reconstructed frame, local intensity maxima were detected, and small regions around these maxima were correlated to a model PSF of our imaging system (note that the same empirical PSF was used over the whole reconstructed volume). The maxima with correlation coefficients above 0.5 were considered as droplets. Localization of these droplets was then further refined using a local quadratic fitting of the intensity maxima, and their positions were stored. A particle tracking algorithm was then used on these positions (simpletracker.m available on mathworks ©Jean-Yves Tinevez, 2019, wrapping matlab munkres algorithm implementation of ©Yi Cao 2009) in order to track the droplets over consecutive frames. A maximal linking distance of 0.5 mm was selected, which corresponds to a maximum particle velocity of 50 $mm.s^{-1}$.

**Fluence estimation**

The amplitude of the OA signal generated by a droplet is proportional to the amount of energy absorbed times the local flight fluence. The absorbed energy depends on the volume of the droplet, which changes according to the size distribution. However, given the narrow distribution of the droplet size, the light fluence can be assumed to be proportional to the average intensity of the OA signals reconstructed for droplets at a given location. For fluence estimation, 50 droplets were selected in a cross-sectional image of the brain. The intensities of the differential OA images for the positions of the selected droplets were fitted to an exponentially decaying function. Specifically, a function of the form (a/z)exp(-bz) was used for least square fitting, being z=sqrt((x-x0)^2+(y-y0)^2)). Curve fitting resulted in optimum values of the parameters a, b, x0 and y0.

**Acknowledgements**

X. L. D. B. acknowledges support from the Werner und Hedy Berger-Janser Stiftung (Application No 08/2019) and the Helmut Horten Stiftung (Project Deep Skin). D. R. acknowledges support from the European Research Council under grant agreement ERC-CoG_2015_682379. X. L. D. B would like to acknowledge Michael Hagander (Microcaps AG, Zürich, Switzerland) for helpful discussions on microdroplet generation.

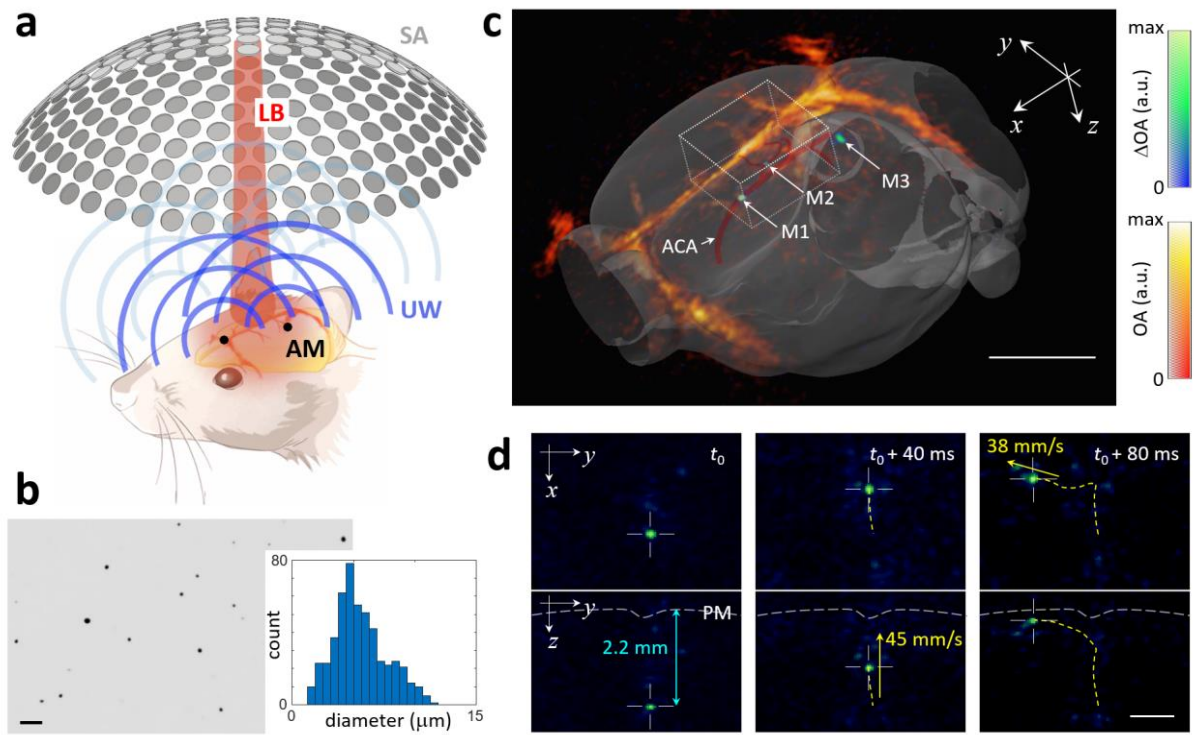

**Figure 1.** Detection and tracking of individual microdroplets with 3D optoacoustic (OA) tomography. (a) Lay-out of the suggested approach. AM – absorbing microdroplets, LB – laser beam, UW – ultrasound waves, SA – spherical array. (b) Bright field (5X) optical microscopy image of the dichloromethane (DCM) microdroplets along with a histogram of the measured diameters (500). Scalebar – 50 μm. (c) OA images of the mouse brain superimposed to the differential (background substracted) OA (ΔOA) images. Three microdroplets in the azygos pericallosal artery (M1), middle internal frontal artery penetrating branches (M2) and pial surface (M3) are labelled in the differential image. The co-registered Allen mouse brain atlas is also shown. ACA – anterior cerebral artery. Scalebar – 3 mm. (d) Maximum intensity projections (MIPs) of the differential OA image along the z and x directions for the rectangular box indicated in (c). Three representative time points are shown. The depth from the pia matter (PM) and the calculated velocities are indicated in blue and yellow, respectively. Scalebar – 1 mm.

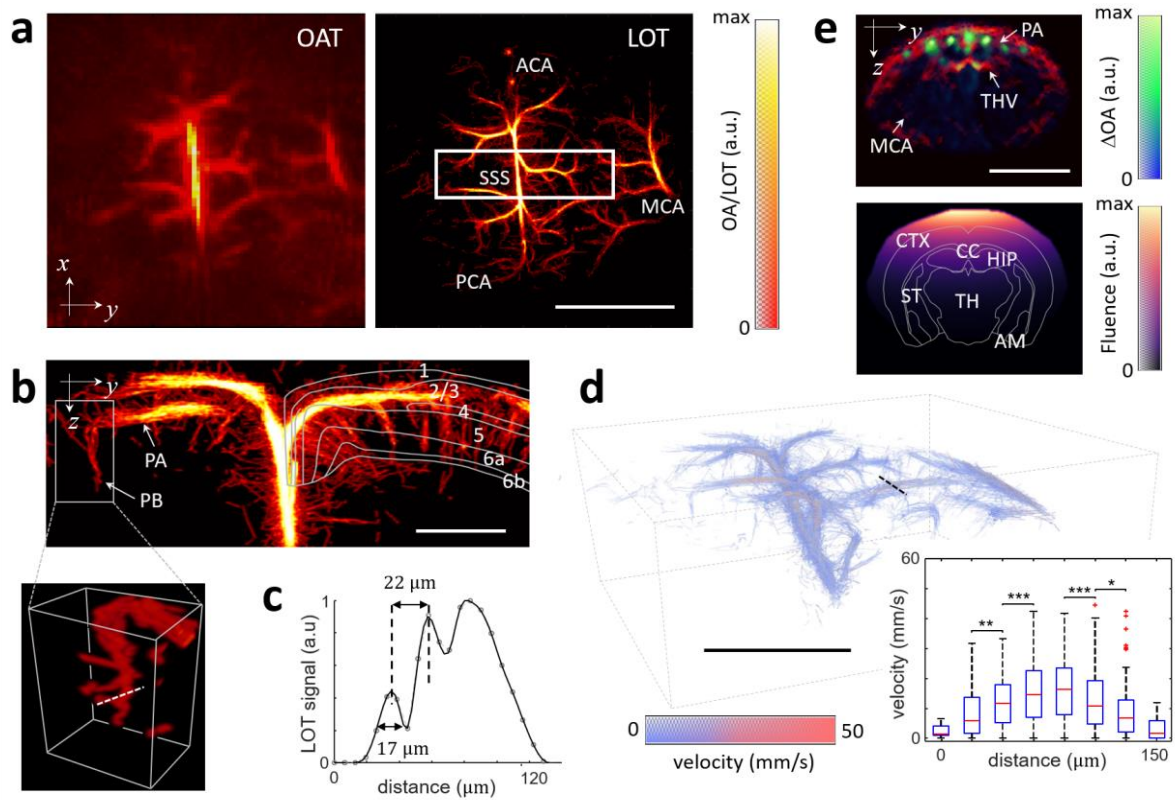

**Figure 2**. Optoacoustic (OA), blood flow and light fluence images. (a) Maximum intensity projections (MIPs) along the z direction (dorsal view) of the 3D images rendered with standard OA tomography (OAT, left) and localization optoacoustic tomography (LOT, right). ACA – anterior cerebral artery, MCA – middle cerebral artery, PCA – posterior cerebral artery, SSS – superior sagittal sinus. Scalebar – 3 mm. (b) MIP along the x direction (coronal view, Bregma -0.22mm to -1.22 mm) of the rectangular box of the LOT image indicated in (a). Cortical layers are indicated. The inlet displays a 3D view of the indicated region. PA – pial arteriole, PB – penetrating branches. Scalebar – 1 mm. (c) Profile of the LOT image for the dashed white line in (b). (d) 3D view of the blood flow image. Inlet shows boxplots of the measured velocities along the dashed black line. Statistical significance is shown. Scalebar - 3 mm. (e) Cross-sectional coronal view (top, approx. Bregma -1.46 mm) of the OA image along with the superposition of 8 differential OA images (ΔOA). The light fluence image (bottom) is derived by fitting the signals of 50 microdroplets in the differential OA images to an exponentially decaying function. MCA – middle cerebral artery, PA – pial artery, THV – transverse hippocampal vein, CTX – cortex, CC – corpus callosum, HIP – hippocampus, TH – thalamus, ST – striatum, AM – amygdala. Scalebar – 5 mm.